\documentstyle[11pt,takashi]{article}
\def \bracket<#1>{\mbox{$\langle {#1}\rangle$}}
\def \brav #1|{\mbox{$\langle {#1}|$}}
\def \cg(#1,#2,#3,#4,#5,#6){\mbox{$(#1,#2,#3,#4|#5,#6)$}}
\def \comm[#1,#2]{\mbox{$\left[{#1},{#2}\right]$}}
\def \ddt #1{\ifmmode{{\partial #1\over\partial t}} \else{${\partial #1\over\partial t}$}\fi}

\def \dotp(#1.#2){\mbox{$(#1\cdot #2)$}}

\def \ketv #1>{\mbox{$|{#1}\rangle$}} 
\def \mate<#1|#2|#3>{\mbox{$\langle {#1}|\,{#2}\,|{#3}\rangle$}}
\def \mated<#1||#2||#3>{\mbox{$\langle {#1}||\,{#2}\,||{#3}\rangle$}}
\def \mvec #1{\mbox{\boldmath{${#1}$}}}
\def \rtov(#1/#2){{\ifmmode{{\sqrt{\frac{#1}{#2}}}} \else{{$\sqrt{\frac{#1}{#2}}$}}\fi}}
\def \sixj(#1,#2,#3,#4,#5,#6){\mbox{$\left\{\matrix {#1&#2&#3\cr#4&#5&#6\cr}\right\}$}}
\def \Trace[#1]{\mbox{${\hbox{Tr} \left\{#1\right\}}$}}
\def \xp(#1.#2){\mbox{${(#1\times #2)}$}}

\def \bra{\mbox{$\langle$}}

\def \etal{{\it et al.}}

\def \ket{\mbox{$\rangle$}}

\def \rhat{\ifmmode{\hat{\mvec r}}\else{$\hat{\mvec r}$}\fi}

\def \ddt #1{\ifmmode{{\partial #1\over\partial t}} \else{${\partial #1\over\partial t}$}\fi}
\def \doublet(#1,#2){{\mbox{$\left( {\matrix{#1\cr#2\cr}} \right) $}}}
\def \triplet(#1,#2,#3){{\left(\matrix{#1\cr#2\cr#3\cr}\right)}}
\def \sixj(#1,#2,#3,#4,#5,#6)
{{\mbox{$\left\{\matrix{#1&#2&#3\cr#4&#5&#6\cr}\right\}$}}}

\def \ninej(#1,#2,#3,#4,#5,#6,#7,#8,#9)
    {\mbox{$\left\{\matrix {#1&#2&#3\cr#4&#5&#6\cr#7&#8&#9\cr}\right\}$}}

\newcommand{\disint}{{\displaystyle \int}}

\newcommand{\mpi}{\tilde{m}_\pi}

\input{epsf.tex}
\begin{document}
\final
\date{}
\title{ Non-mesonic Weak Decays of Light Hypernuclei \\
        in the Direct Quark and the One-Pion Exchange Mechanisms}
\author{
T.~Inoue$^{(a)}$\thanks{e-mail: inoue@nt.phys.s.u-tokyo.ac.jp},
M.~Oka$^{(b)}$, T.~Motoba$^{(c)}$ and K.~Itonaga$^{(d)}$ \\
{\sl $^{(a)}$ Department of Physics, University of Tokyo,}
{\sl Bunkyo, Tokyo 112 Japan}  \\
{\sl $^{(b)}$ Department of Physics, Tokyo Institute of Technology,}
{\sl Meguro, Tokyo 152 Japan}  \\
{\sl $^{(c)}$ Laboratory of Physics, Osaka Electro-Communication
University,}\\
{\sl Neyagawa, Osaka 572 Japan} \\
{\sl $^{(d)}$ Laboratory of Physics, Miyazaki Medical College,}
{\sl Miyazaki 889-16 Japan}
}
\maketitle
\abstract{
 Contributions of the direct quark mechanism are studied in
 the nonmesonic weak decays of light hypernuclei.
 The $\Lambda N \to NN$ transition is described
 by the one pion exchange mechanism and the direct quark mechanism,
 induced by the four-quark vertices in the effective weak Hamiltonian.
 By employing a realistic wave function of the $\Lambda$
 inside the hypernuclei, nonmesonic decay rates of
 $^4_{\Lambda}$H, $^4_{\Lambda}$He, and  $^5_{\Lambda}$He
 are calculated.
 The results show that the direct quark mechanism
 is significantly large  and gives a large $\Delta I =3/2$ contribution
 in the $J=0$ channel.
The relative phase between the one-pion exchange
and the direct-quark contributions is determined so that the
effective weak Hamiltonian for quarks give both of them consistently.
We find that the sum of these two contributions reproduce
the current available experimental data fairly well.
         }
\newpage
 \section{Introduction}%-------------------------------------

An $S=-1$ hypernucleus decays from its ground state to non-strange
hadrons,
emitting roughly 170 MeV ($\sim M_{\Lambda}- M_N$) of extra energy
\cite{Choen}.
As is the decay of free $\Lambda$, it can emit a pion + 30 MeV.
It is, however, known that the pionic decay is strongly suppressed
because
the remaining energy is too small for the produced nucleon to go above
the Fermi surface of the nucleus.
Consequently, the main decay mode of (heavy) hypernuclei is the decay
without a pion emission, called nonmesonic decay.
The nonmesonic decay releases the full 170 MeV of energy and therefore
is not
Pauli blocked.  The final state is dominated by two nucleons ejected
in the opposite directions so that the momentum is conserved,
although three-body processes may not be negligible \cite{threebody1}.
A natural way of describing the nonmesonic decay is to assume
that the emitted (virtual) pion from the $\Lambda$ decay is absorbed
by a nucleon in the nucleus.
This is called the one-pion exchange (OPE) weak transition and
has been studied by many authors \cite{BD,TTB,MG,Ramos,Fujii,BMZ}.

On the other hand, the final $NN$ state has a large relative momentum so
that
the short-range interactions are also important \cite{CHK,Shmatikov}.
Two of the present authors have proposed that the quark structure
of the baryons provides a new mechanism for the nonmesonic
$\Lambda N \to NN$, called direct quark (DQ) mechanism
\cite{OkaInoue,InoueOka,OkaInoue2,InoueTakeuchiOka}.
We calculated transition amplitudes for the S wave initial
states in both OPE and DQ,
and found that the contribution of the DQ is in general
as large as that of the OPE.
DQ is even dominant in the $L=0$ to $L=0$ transitions.
It gives a large contribution in the J=0 (L=0,S=0) transitions
and thus contributes to the nn (I=1) final state.
We found that the $nn/pn$ ratio of the final state becomes larger
than that predicted in OPE.
It was also found that the $J=0$ transition amplitudes have
a significant $\Delta I =3/2$ contribution.

In this paper, we apply the DQ transition potential to
the decay of three light hypernuclei,
$^4_{\Lambda}$H, $^4_{\Lambda}$He,  and $^5_{\Lambda}$He in order to
study how the features of DQ are reflected in the decay observables of
simple hypernuclei.
We employ realistic wave functions for the hypernuclei,
and evaluate the matrix elements of the two-body weak transition
potentials
in OPE and DQ, and also of the potential
given by the superposition of OPE and DQ.
The relative phase of OPE and DQ are determined so that the effective
weak Hamiltonian for quarks give both consistently.
The results are compared with current available data.
We find that the OPE+DQ superposed transition potential reproduce
the currently available experimental data fairly well.

In sec.~2, we present the two-body weak transition potentials for
$\Lambda N \to NN$ in DQ and OPE.
In sec.~3, the nuclear wave functions are given.
In sec.~4, the results of our calculation are presented and are compared
with experiment.
Sec.~5 is devoted to conclusions.

 \section{ Weak transition potential for $\Lambda N \to NN$ }%-----------

\begin{figure}[t]
   \epsfysize = 40 mm
   \centerline{ \epsfbox{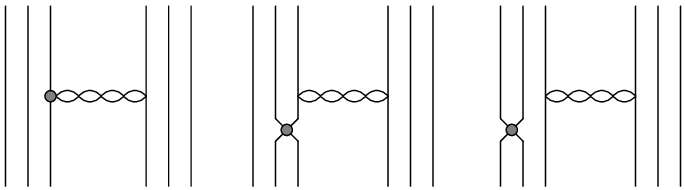} }
\caption{The quark diagrams for the $\Lambda N \to NN$ transition.
         $\bullet$ denotes a weak 4-quark vertex.}
\label{fig:fig1}
\end{figure}
\begin{figure}[t]
  \epsfysize = 40 mm
  \centerline{
    \epsfbox{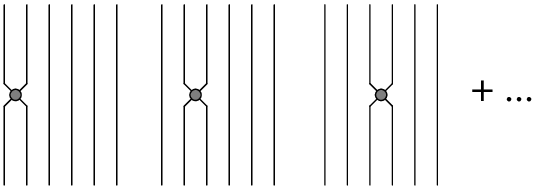}
              }
\caption{The quark diagrams for the $\Lambda N \to NN$ transition.
         $\bullet$ denotes a weak 4-quark vertex.}
\label{fig:fig2}
\end{figure}

\begin{figure}[t]
   \epsfysize = 50 mm
    \centerline{
      \epsfbox{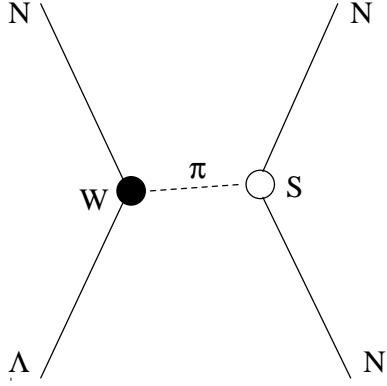}
               }
 \caption{One pion exchange diagram.}
 \label{fig:opeex}
\end{figure}

  Considering that a baryon has three constituent quarks,
  $\Lambda N \to NN $ process can be described by the diagram,
  such as those shown in Fig. \ref{fig:fig1} and Fig. \ref{fig:fig2}.
  In the diagrams shown in Fig. \ref{fig:fig1},
  the strangeness changing weak interaction
  and an emission of a constituent quark anti-quark pair
  take place in the $\Lambda$ hyperon, and the pair is absorbed
  by a constituent quark in the nucleon.
  The diagrams in Fig. \ref{fig:fig2} show the
  weak interaction of two constituent quarks in a totally
  anti-symmetric six constituent quark state.
  The first picture might be well represented by the diagram in
  Fig. \ref{fig:opeex},
  where the baryon is a Dirac particle and couples to, for example,
  a pion by a phenomenological Yukawa type vertex.
  This is the one called the meson exchange mechanism
  and has been studied well  \cite{BD,TTB,MG,Ramos,Fujii,BMZ}.
  Though this picture is very natural,
  one sees that this picture cannot be valid in the region
  where the two baryons overlap with each other.
  In such a region, the diagrams in Fig. \ref{fig:fig2},
  namely the direct quark process,
  might cause the $ \Lambda N \to NN $ transition.

Recently, Inoue, Takeuchi and Oka pointed out the importance of
the direct quark (DQ) processes
and derived the DQ transition potential for the $\Lambda N \to NN$
systems\cite{OkaInoue,InoueOka,OkaInoue2,InoueTakeuchiOka}.
There we employed the nonrelativistic constituent quark model for the
baryons and the two baryon states are constructed according to the
quark cluster model, which takes the full quark antisymmetrization
into account,
\eq
    \ketv B_1 B_2 (k,L,S,J) >
   = {\cal A} \ketv \phi_1 \phi_2 \chi(k) (L,S,J) >
 \label{eqn:qcmw}
\eq
where $\phi_1$ and $\phi_2$ denote the internal wave functions of
the baryon $B_1$ and $B_2$ and $\chi(k)$ describes the
radial part of the relative motion of the baryons with $k$ relative
momentum.
${\cal A}$ is the quark antisymmetrization operator.

\begin{figure}[t]
   \epsfysize = 120 mm
   \centerline{
   \epsfbox{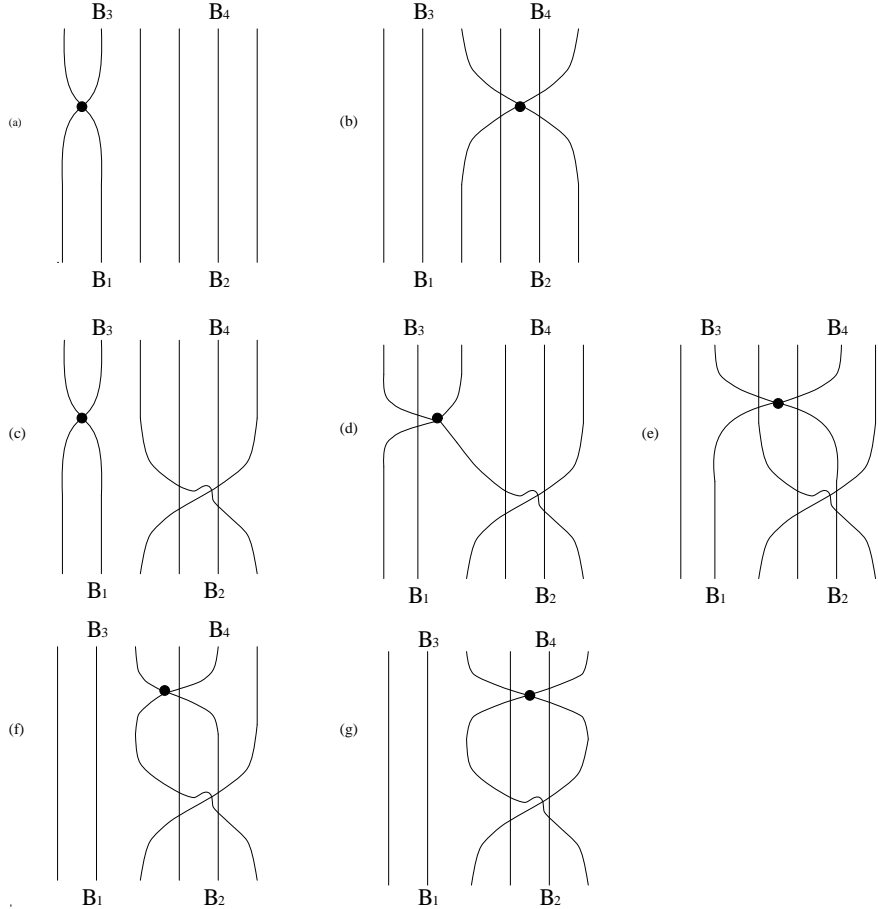}
               }
\caption{The direct quark transitions.
         $\bullet$ denotes a weak 4-quark vertex.}
 \label{fig:seventerm}
\end{figure}

The weak interaction is represented by the effective Hamiltonian,
\eq
% H_{eff}^{\Delta S=1}\left(\mu=\mu_0 \right)=
   H_{eff}^{\Delta S=1} =
  -\frac{G_f}{\sqrt 2}\sum_{r=1,r\ne 4}^6K_r O_r
\label{eqn:heff}
\ ,
\eq
where
\eq\begin{array}{c|c|c|c|c}
  \quad  K_1   \quad  &   \quad  K_2   \quad
  &
  \quad  K_3   \quad  &   \quad  K_5   \quad
  &
  \quad  K_6   \quad \\
 \hline
   -0.284 &  0.009  & 0.026 & 0.004 & -0.021 \\
\end{array}
\nonumber
\eq
and
\eq
  O_1 &=& (\bar d_{\alpha}s_{\alpha})_{V-A}
          (\bar u_{\beta}u_{\beta})_{V-A}
         -(\bar u_{\alpha}s_{\alpha})_{V-A}
           (\bar d_{\beta}u_{\beta})_{V-A}
  \\
  O_2 &=& (\bar d_{\alpha}s_{\alpha})_{V-A}
           (\bar u_{\beta}u_{\beta})_{V-A}
         +(\bar u_{\alpha}s_{\alpha})_{V-A}
           (\bar d_{\beta}u_{\beta})_{V-A}\nonumber
  \\
     & &+2(\bar d_{\alpha}s_{\alpha})_{V-A}
         (\bar d_{\beta}d_{\beta})_{V-A}
        +2(\bar d_{\alpha}s_{\alpha})_{V-A}
         (\bar s_{\beta}s_{\beta})_{V-A}
  \\
  O_3 &=& 2(\bar d_{\alpha}s_{\alpha})_{V-A}
           (\bar u_{\beta}u_{\beta})_{V-A}
         +2(\bar u_{\alpha}s_{\alpha})_{V-A}
           (\bar d_{\beta}u_{\beta})_{V-A}\nonumber
  \\
     & &-(\bar d_{\alpha}s_{\alpha})_{V-A}
         (\bar d_{\beta}d_{\beta})_{V-A}
        -(\bar d_{\alpha}s_{\alpha})_{V-A}
         (\bar s_{\beta}s_{\beta})_{V-A}
  \\
  O_5 &=& (\bar d_{\alpha}s_{\alpha})_{V-A}
       (\bar u_{\beta}u_{\beta}+\bar d_{\beta}d_{\beta}
      + \bar s_{\beta}s_{\beta})_{V+A}
  \\
  O_6 &=& (\bar d_{\alpha}s_{\beta})_{V-A}
       (\bar u_{\beta}u_{\alpha}+\bar d_{\beta}d_{\alpha}
      + \bar s_{\beta}s_{\alpha})_{V+A}
\ .
\eq
This is derived from the $W$ exchange diagrams in the standard theory,
by taking  perturbative corrections
due to the strong interaction \cite{Paschos}.
We evaluated the two-baryon matrix elements of the effective
weak Hamiltonian $H_{eff}^{\Delta S=1}$,
using the six-quark wave functions given in eq.(\ref{eqn:qcmw}),
\eq
    V(k,k')_{{L_i,S_i,J}\atop{L_f,S_f,J}}=
      \mate<NN(k',L_f,S_f,J)|H_{eff}^{\Delta S=1}|
             \Lambda N(k,L_i,S_i,J)> \ .
\label{eqn:defofv}
\eq
Here, $k$, $L$, $S$, $J$ stand for the relative momentum of two baryons,
the orbital angular momentum, the spin, and the total angular momentum
of the initial
and the final states, respectively.
The obtained two-body transition amplitudes are regarded as the
transition
potential in the momentum representation
for each channel specified by $L$, $S$ and $J$.
Fig. \ref{fig:seventerm} shows various diagrams contributing
to the transition potential, where the dots represent
the weak four quark vertices.
The explicit forms of the transition potential
is given in ref. \cite{InoueTakeuchiOka}.

In the present study, we calculate the decays of the s-shell hypernuclei
from its ground states.  If we employ the simplest shell model wave
functions
for the hypernuclei, we only need the $L_i=0$ transition potentials.
The relevant transition channels are given in Table 1.
 The direct quark induced transition potential
 depends on two quark model parameters that
 are the constituent quark mass, $m$, and the Gaussian parameter, $b$.
 We use $m=313=M_N/3$ MeV and $b=0.5$ fm in the present calculation.

As for the meson exchange weak transition, we here consider only the
one-pion
exchange (OPE) process, and will discuss other contributions later.
The pion exchange transition potential is obtained by evaluating the
diagram
like Fig. \ref{fig:opeex}.
We here employ the following strong and weak pion vertices,
\eq
    H_{NN\pi}^s
       &=&  i g_s  \bar\psi_p\gamma_5\pi^0\psi_p
         -  i g_s \bar\psi_n\gamma_5\pi^0\psi_n
         + i \sqrt2  g_s\bar\psi_p\gamma_5\pi^+\psi_n
         + i \sqrt2  g_s\bar\psi_n\gamma_5\pi^-\psi_p
       \label{eqn:phenostrong}
 \\
   H_{\Lambda N \pi}^w
        &=&
          i g_w \bar\psi_n(1+\lambda \gamma_5)\pi^0 \psi_{\Lambda}
    - i \sqrt2 g_w \bar\psi_p(1+\lambda \gamma_5)\pi^+ \psi_{\Lambda}
       \label{eqn:phenoweak}
\eq
where the strong coupling constant $g_s$ is chosen as the standard value
of the
$\pi N N$ coupling:  $g_s = - 13.26$.
The weak coupling constants $g_w$ and $\lambda$ are determined so as to
reproduce the $\Lambda \to N \pi$ decay amplitudes:
\eq
  g_w     &=&   - 2.3 \times 10^{-7}
  \\
  \lambda &=&  -6.9
\ .
\eq

The relative sign of $g_s$ and $g_w$ is important when we
consider the interference of OPE and DQ contributions.
Namely, $g_w$ must be chosen consistently with our weak
Hamiltonian for quarks and the baryon wave functions.
In order to check the consistency,
we use the soft pion theorem.
%%%
Consider the parity-violating part of the $\Lambda\to n\pi^0$
decay matrix element in the soft pion limit,
\eq
  \lim_{q\to 0} \mate< n\pi^0(q)|H_{eff}(PV)| \Lambda>
   = - {i\over f_{\pi}} \mate< n|[Q_5^3, H_{eff}(PV)]| \Lambda>
\ .
\eq
The soft-pion limit is taken by assuming 
\eq
   \mate< \pi^0(q)|A_{\mu}^a (x)| 0> = -i f_{\pi} q_{\mu} 
   \exp(iq\cdot x)
\ .
\eq
Note that the Goldberger Treiman relation 
\eq
    f_{\pi} = \frac{M g_A }{(- g_s)}
\eq
is satisfied, where $g_s$ is the strong $\pi NN$ coupling constant
defined in eq.(\ref{eqn:phenostrong}).
Because $H_{eff}(\Delta S = 1)$ consists only of the left-handed
currents and the flavor singlet right-handed currents, we have
\eq
  [Q_5^3, H_{eff}(PV)] = [Q_R^3- Q_L^3,  H_{eff}(PV)] \\
  = -[Q_R^3+ Q_L^3,  H_{eff}(PC)] = -[I_3,  H_{eff}(PC)] 
\eq
Thus we obtain
\eq
   \lim_{q\to 0} \mate< n\pi^0(q)|H_{eff}(PV)| \Lambda>
   =  {i g_s\over 2M g_A} \mate< n|H_{eff}(PC)| \Lambda>
\eq
Comparing this with the effective hamiltonian eq.(\ref{eqn:phenoweak}),
the weak $\pi NN$ coupling constant is given by
\eq
  g_w =  {g_s\over 2M g_A} \mate< n|H_{eff}(PC)| \Lambda> < 0
\eq 
after evaluating the matrix element in the right hand side, 
which is positive
with our definition of the Hamiltonian $H_{eff}^{\Delta S=1}$
and the quark model wave function.
%%%

\begin{table}[t]
\caption{ One pion exchange induced transition potential}
\label{tbl:opepot}
\begin{center}
\begin{tabular}{ccrccc}
           &                   & spin-orbital              & $I_{NN}$
   &       & Potential \\
   \hline
   $a_p$   & $p\Lambda \to pn$ & ${}^{1}S_0 \to {}^{1}S_0$ & 1
   &       & ${  \displaystyle
                -\frac{1}{\sqrt2}\lambda \frac{\mpi}{2\bar M}f(r)
             }$
   \\
   $b_p$   &                   &           $\to {}^{3}P_0$ & 1
   &       & ${ \displaystyle
                  i \frac{1}{\sqrt2} V(r) f(r)
             }$
   \\
   $c_p$   &                   &  ${}^{3}S_1\to {}^{3}S_1$ & 0
   &       & ${ \displaystyle
                -\frac{1}{\sqrt2}\lambda \frac{\mpi}{2\bar M}f(r)
             }$
   \\
   $d_p$   &                   &           $\to {}^{3}D_1$ & 0
   &       & ${ \displaystyle
                -6\lambda \frac{\mpi}{2\bar M}T(r)f(r)
             }$
   \\
   $e_p$   &                   &           $\to {}^{1}P_1$ & 0
   &       & ${ \displaystyle
                 i \sqrt{\frac32}V(r)f(r)
             }$
   \\
   $f_p$   &                   &           $\to {}^{3}P_1$ & 1
   &       & ${\displaystyle  
                - i \sqrt{\frac13} V(r)f(r)
             }$
   \\
   $a_n$   & $n\Lambda\to nn$  &  ${}^{1}S_0\to {}^{1}S_0$ & 1
   &       & ${ \displaystyle
               -\lambda \frac{\mpi}{2\bar M}f(r)
             }$
   \\
   $b_n$   &                   &           $\to {}^{3}P_0$ & 1
   &       & ${ \displaystyle  
                i V(r)f(r)
             }$
   \\
   $f_n$   &                   &  ${}^{3}S_1\to {}^{3}P_1$ & 1
   &       & ${\displaystyle
                - i \sqrt{\frac23} V(r)f(r)
             }$
\end{tabular}
\end{center}
\end{table}

  In Table \ref{tbl:opepot},
  we summarize the radial part of the OPE transition potential
  for the nine possible channels which appear in the calculation for the
  s-shell hypernuclei.
  The function in the table are defined by
\eq
   f(r)&=&
    \frac{g_w g_s}{4\pi}\frac{\mpi}{2\bar M} \mpi
       \frac{e^{-\mpi r}}{\mpi r}
   \\
   V(r)&=& 1 + \frac{1}{\mpi r}
   \\
   T(r)&=&\frac13 + \frac{1}{\mpi r} + \frac{1}{\mpi^2 r^2}
\eq
 when the form factor is not taken account.
 Here $\bar M$ is the average mass of the baryons
 and $\tilde{m}_{\pi} = \sqrt{m_{\pi}^2-q_0^2}$ is an effective pion
mass
 introduced in order to take care of the finite energy transfer.
The radial functions are modified when the form factor with a cutoff
$\Lambda_{\pi}$ is introduced, so that
\eq
    f(r)      &\to&
      f(r) -  \left( \frac{\Lambda_{\pi}}{\mpi} \right)^3
              f(\Lambda_{\pi} r)
    \\
    f(r)V(r)  &\to&
      f(r)V(r) -  \left( \frac{\Lambda_{\pi}}{\mpi} \right)^3
               f(\Lambda_{\pi} r) V(\Lambda_{\pi} r)
    \\
    f(r)T(r)  &\to&
      f(r)T(r) -  \left( \frac{\Lambda_{\pi}}{\mpi} \right)^3
               f(\Lambda_{\pi} r) T(\Lambda_{\pi} r)
\ .
\eq
In the present calculation, we choose $\Lambda_{\pi}^2 = 20 \tilde
m_{\pi}^2$,
according to ref. \cite{TTB}.

We regard DQ and OPE as independent of each other.
Fig. \ref{fig:seventerm} shows the various
diagrams which contribute to the DQ transition potential.
Among them, the diagrams (b) - (g) contain the quark exchange between
the two baryons, and therefore the range of the DQ potential is
determined
by the size of the quark wave function of the baryon.
Thus, their contributions are independent from the one-pion exchange,
as the strong repulsion due to the quark-exchange is independent
from the one-pion exchange potential in the nuclear force.

As the diagram (a), on the other hand, gives a long-range contribution,
it may cause a double counting problem.
This is the process in which $\Lambda$ decays into $n$ with
another nucleon as a spectator.  Its matrix element is nonzero
only through high momentum components of the two baryon
wave functions and therefore comes mainly from the short-range
correlation of the baryons.
Because the one-pion exchange seems not responsible for the
short-range correlation, we also regards the diagram (a) as independent.
It should also be noted that the contribution of the diagram (a)
is in general small as is expected.

 \section{ Initial and final wave functions }%----------------

For the s-shell hypernuclei, we employ a shell model wave function
for the nucleon core, denoted by cluster, $x$.
It has been pointed out that it is important to solve
the $\Lambda$ single particle state carefully using the realistic
$\Lambda$-nucleon interaction{\cite{motoba547}}.
 We denote the relative motion of the $\Lambda$ hyperon
 against the nucleon cluster $x$ by a wave function
 $\Psi( \mvec r_{\Lambda} )$  with $\mvec r_{\Lambda}$,
 relative coordinate.
 The Hamiltonian $H$ of the $\Lambda$-$x$ system
 is given by the sum of the kinetic energy $T$ and
 a potential energy $V_{\Lambda x}$.
%\eq
%     H =  T + V_{\Lambda x}
%  \ .
%\eq
 The potential energy $V_{\Lambda x}$ is obtained by folding
 the potential energy between $\Lambda$ and the nucleons.
 In the present study, we construct it by folding the YNG potential
 with $(0s)^x$ where $x$ stands for the number of nucleon in the cluster
$x$.
 The YNG potential is a three-range Gaussian potential
 which reproduces the G-matrix for the Nijmegen
 $\Lambda N$ potential model D \cite{Yamamoto}.
 The Fermi momentum $k_f$ is fixed $0.9$ fm$^{-1}$
 for light hypernuclei.

 The Schr\"odinger equation is solved
 variationally using the local Gaussian basis functions.
 We expand the wave function for the state which has
 orbital angular momentum $l$ and the spin $S$ as
\eq
  \Psi(\mvec r_{\Lambda}) = \sum_{d} f_{lj}(d) \ketv \Phi(l;d), S ; j >
  \label{eqn:baseexp}
 \ .
\eq
 where $\mvec j = \mvec l + \mvec S$.
 The spin of hypernuclei, $\mvec S$, is the sum of total angular
 momentum of the cluster $x$, $\mvec{J_x}$,
 and the spin of $\Lambda$, $\mvec{S_{\Lambda}}$.
 In eq.(\ref{eqn:baseexp}), $\Phi(l;d)$ is the basis function given by
\eq
  \Phi(l;d)
  &=& \phi_l(r_{\Lambda};d)Y_l(\hat{\mvec r}_{\Lambda})
  \\
  \phi_l(r_{\Lambda};d) &=& 4 \pi (\sqrt{\pi} b_{\Lambda x})^{-3/2}
               \exp[ \frac{-(r_{\Lambda}^2+d^2)}{2 b_{\Lambda x}} ]
               {\cal J}_l( \frac{r_{\Lambda} d}{b_{\Lambda x}^2} )
  \\
  b_{\Lambda x} &=& \sqrt{ \frac{(M_{\Lambda}+ x M_N)}
                                { x M_{\Lambda} } } b_N
 \label{eqn:blx}
 \ .
\eq
 The parameter $d$ can be considered as a generator coordinate
describing
 the distance between two clusters.
 The ${\cal J}_l$ is the $l$-th order modified spherical Bessel
function.
 The choice of $b_{\Lambda x}$ as in eq(\ref{eqn:blx}) enables us to use
the
 Talmi-Moshinsky transformation coefficients
 in calculating the two-body matrix elements later.
 The variational amplitudes $f_{l j}$ satisfy the generator coordinate
method
 equation
\eq
  \sum_{d_2}\left[
       H_{l j}(d_1,d_2) - E N_{lj}(d_1,d_2)
            \right] f_{l j}(d_2) =0
\label{eqn:GCM}
\eq
 where the energy and the normalization kernels are given by
\eq
 H_{l j}(d_1,d_2) &=&
 \mate< \Phi(l;d_1), S ; j | H | \Phi(l;d_2), S ; j >
 \\
 N_{lj}(d_1,d_2)  &=&
 \mate< \Phi(l;d_1), S ; j | 1 | \Phi(l;d_2), S ; j >
 \ .
\eq

\begin{figure}[t]
   \epsfysize = 80 mm
   \centerline{ \epsfbox{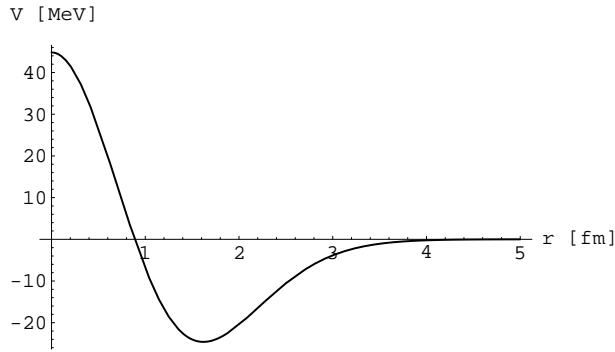} }
  \caption{ Folded potential for $^5_{\Lambda}\mbox{He}$ for even $l$
state. }
\label{fig:foldpot}
\end{figure}

 For the hypernucleus $^5_{\Lambda}\mbox{He}$,
 we take the Gaussian parameter $b_N$ for $0s$ as 1.358 fm
 \cite{Motoba,Itonaga}.
 We present the folding potential for even $l$ state
 in Fig \ref{fig:foldpot}.
 With the following seven values for d,
\eq
    d = 0.0, 1.0, 2.5, 4.0, 5.5, 7.0 \mbox{ and } 8.0 \mbox{ fm}.
\eq
 we obtain the ground state energy  $E=-3.08$ MeV
 and corresponding variational amplitudes.

\begin{figure}[t]
   \epsfysize = 80 mm
   \centerline{ \epsfbox{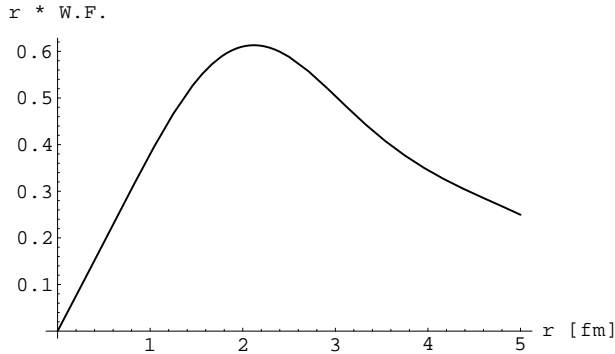} }
   \caption{ Wave function for $\Lambda$-$\alpha$ system. }
\label{fig:wave}
\end{figure}

 Fig \ref{fig:wave} shows the obtained wave function $\Psi( \mvec r )$.
 This wave function is considerably different from the one which
 is obtained with the one range Gaussian (ORG) potential.
 As shown in Fig \ref{fig:foldpot}, the potential
 $V_{\Lambda x}$ is repulsive at the short distance,
 reflecting the presence of a repulsive soft core in the YNG
interaction.
 Therefore the obtained wave function is pushed out.
 This feature gives rise to a sizable difference in
 the r.m.s. distance between $\alpha$ and $\Lambda$,
 3.06 fm for YNG while 2.69 fm for ORG \cite{Motoba,Itonaga}.
 Motoba \etal~ applied this wave function to the pionic decay
 of the s-shell hypernuclei and found that the YNG
 wave function gives a better agreement to experiment
 than the ORG one \cite{motoba547,Kumagai-Fuse}.

 The wave function $\Psi( \mvec r_{\Lambda} )$
 for $^4_{\Lambda}\mbox{He}$ (or $^4_{\Lambda}\mbox{H}$)
 is calculated in the same way.
 In the 4-body case, the Gaussian parameter $b_N$
 is taken as 1.65 fm \cite{Motoba,Itonaga}.

 The $\Lambda N$ wave function, $\Psi_{\Lambda N}$, is given by the
 product of the above wave function, $\Psi(\mvec r_{\Lambda})$,
 and  nucleon wave function, $\Psi_{N}(\mvec r_N)$,
 which is assumed as a Gaussian,
\eq
      \Psi_{\Lambda N}( \mvec r_{\Lambda}, \mvec r_N )
   =  \Psi(\mvec r_{\Lambda}) \times \Psi_N(\mvec r_N)
\ .
\eq
 We perform the Talmi-Moshinsky transform for
 the two-baryon $\Lambda N$ wave function
 $\Psi_{\Lambda N}$ in order to separate the
 relative motion and the center of mass motion.
 Defining $ \mvec r = \mvec r_{\Lambda} - \mvec r_{N}$
 and the two-baryon center-of mass coordinate $\mvec R$,
 the product wave function can be expanded as
\eq
   \Psi_{\Lambda N}( \mvec r, \mvec R)
   = \sum_{n N} W_{nN} u_{n 0}(r,b_r)Y_{00}(\hat {\mvec r})
                       u_{N 0}(R,b_R)Y_{00}(\hat {\mvec R})
 \ ,
\eq
 where $u_{n0}$ is the harmonic oscillator eigenfunctions for
$(n,l=0)$,
 and
\eq
 b_r  &=& \sqrt{ \frac{M_{\Lambda}+M_N}
                      {M_{\Lambda}     } } b_N
 \\
 b_R  &=& \sqrt{ \frac{ M_{\Lambda}+ x M_N }
                      { (x-1)(M_{\Lambda}+M_N ) }  } b_N
\ .
\eq
 The b parameters, A=4 are $b_r = 2.24$ fm, $b_R =1.61$ fm,
 and for A=5 are $b_r = 1.84$ fm, $b_R = 1.21$ fm, respectively.

  We consider the short-range correlation between $\Lambda$ and $N$.
  In evaluating the OPE transition amplitude,
  we use the short range correlation obtained
  for the Nijmegen $\Lambda$-$N$ potential model D
\cite{Motoba,Itonaga}.
  The Nijmegen  model D, however, has a hard repulsive core at short
distance,
  which seems inconsistent with the constituent quark picture.
  Therefore, in evaluating the matrix element of the DQ transition
potential,
  we use the following form of the short range correlation,
\eq
   u_{n 0}(r,b_r)  & \rightarrow  &
   \frac{1}{N} u_{n 0}(r,b_r) \left( 1 - C_i \exp[ - \frac{r^2}{r_i^2} ]
\right)
   \ .
\eq
  Here, $C_i$ and $r_i$ are parameters, and $N$ is a normalization
factor.

  As wave function of the outgoing two nucleons in the final state,
  we employ the scattering state obtained in the Nijmegen model D
  \cite{Motoba,Itonaga}.
  On the other hand, we use the plane wave with the following form
  of the short range correlation  in evaluating the DQ contribution
  because of the same reason stated above.
\eq
   j_l(k r)  & \rightarrow &
   j_l(k r) \left( 1 - C_f \exp[ - \frac{r^2}{r_f^2} ] \right)
\eq
 We evaluate the DQ matrix elements for several choices
 of the parameters for the short-range correlation,
 while we assume $C_f=C_i$ and $r_f=r_i$ for simplicity.

\section{ Results and Discussion }
 In this section, we present our results and compare them with
experiment.
 The experimental data are taken from
 ref \cite{Szymanski,KishiWein,Ajimura,Kishimoto,Schu,Schu2,Val,Outa}.

The nonmesonic decay rate of the hypernucleus is given by
\eq
 \Gamma = \sum_f \delta(E.C.) |\langle \Psi_f| V^{weak}|\Psi_i\rangle
|^2
\eq
where $\delta(E.C.)$ stands for the delta function for the energy
conservation.
For the final state, we observe only the two nucleons outgoing in the
decay.
Denoting the momenta of the nucleons as $k_1$ and $k_2$,  we find
\eq
 \Gamma = \int {dk_1\over (2\pi)^3} {dk_2\over (2\pi)^3} 2\pi
\delta(E.C.)
         \sum_{res} |\langle k_1, k_2, \Psi_{res}| V^{weak} |
\Psi_i\rangle |^2
\eq
and
\eq
 \delta(E.C) = \delta \left(
                 M_{res}+2 M_N+ \frac{k_1^2}{2M_N}+ \frac{k_2^2}{2M_N}
               - M_{res}-M_{\Lambda}-E_{\Lambda}-M_N-E_N
                     \right)
\ .
\eq
Here $V^{weak}$ is a two-body transition potential, and
$\Psi_i$ can be expressed as
\eq
  \Psi_i = \sum_{ch} \Psi_{res} \times \Psi_{\Lambda N(ch)}
           \times (\mbox{cfp for initial state for channel $ch$})
\eq
for which the spin is implicit.
%
% so that
%\eq
% \sum_{res} |\langle k_1, k_2, \Psi_{res}| V^{weak} | \Psi_i\rangle |^2
%    =  |\langle k_1, k_2| V^{weak} | \Lambda N\rangle |^2 \times
%   (\mbox{cfp for initial state})^2
%\ .
%\eq
%
It is easy to obtain the (cfp) for the initial state
in the present $(0s)^x + \Lambda$ configuration,
resulting a factor for the number of pairs and their spin
average factor for each channel, labeled by $a$-$f$ in Table 1,
of the $\Lambda N$ states.
It should be noted that these channels do not interfere with others.
Thus we obtain $\Gamma = \sum_{ch} \Gamma_{ch}$, and the decay rate for
each channel is given by
\eq
  \Gamma_{ch}
  &=&
  \mbox{[\# of $\Lambda N$ pairs]} \times \mbox{[spin average factor]}
\times
  \\
  & &
  \int {dk_1\over (2\pi)^3} {dk_2\over (2\pi)^3} 2\pi \delta(E.C.)
  |\langle k_1, k_2| V_{ch}^{weak} | \Lambda N\rangle |^2
\ .
\eq

\subsection{ $^5_{\Lambda}\mbox{He}$ }

  In $^5_{\Lambda}\mbox{He}$, there are two
  $\Lambda p$ bonds and two $\Lambda n$ bonds.
  The $\Lambda p$ or $\Lambda n$ system takes
  both spin S=0 and spin S=1 state.
  Thus the spin average factor becomes $1/4$ ($3/4$)
  for $S=0$ ($S=1$) channel.
  The total nonmesonic decay rate of $^5_{\Lambda}\mbox{He}$
  is given by
\eq
 \Gamma_{nm}(^5_{\Lambda}\mbox{He})
      &=&
          \Gamma_a^p + \Gamma_b^p
       +  \Gamma_c^p + \Gamma_d^p + \Gamma_e^p + \Gamma_f^p
       +  \Gamma_a^n + \Gamma_b^n + \Gamma_f^n
\eq
  where $\Gamma_a^p \sim \Gamma_f^n$ are the partial decay rates
  and calculated by
\eq
   \Gamma_{a,b}^p
    &=& 2 \frac14
   \disint \frac{d^3 \mvec k}{(2\pi)^3}
   \disint \frac{d^3 \mvec K}{(2\pi)^3}
   (2 \pi)\delta(E.C.)
   \left|
      \bra {NN} |V_{a,b}^p| {\Lambda N} \ket
   \right|^2
  \\
   \Gamma_{c,d,e,f}^p
   &=& 2  \frac34
   \disint \frac{d^3 \mvec k}{(2\pi)^3}
   \disint \frac{d^3 \mvec K}{(2\pi)^3}
   (2 \pi)\delta(E.C.)
   \left|
      \bra {NN} |V_{c,d,e,f}^p|{\Lambda N} \ket
   \right|^2
   \\
   \Gamma_{a,b}^n
   &=& 2 \frac14
   \disint \frac{d^3 \mvec k}{(2\pi)^3}
   \disint \frac{d^3 \mvec K}{(2\pi)^3}
   (2 \pi)\delta(E.C.)
   \left|
      \bra {NN} |V_{a,b}^n|{\Lambda N} \ket
   \right|^2
  \\
  \Gamma_{f}^n
   &=& 2  \frac34
   \disint \frac{d^3 \mvec k}{(2\pi)^3}
   \disint \frac{d^3 \mvec K}{(2\pi)^3}
   (2 \pi)\delta(E.C.)
   \left|
      \bra {NN} |V_{f}^n|{\Lambda N} \ket
   \right|^2
 \ .
\eq

\begin{table}[t]
\caption{ Calculated nonmesonic decay rates of $^5_{\Lambda}\mbox{He}$
          in the one pion exchange mechanism
          (in the unit of $\Gamma_{\Lambda}$,
            the decay rate of the free $\Lambda$). }
\label{tab:operesults}
\begin{center}
\begin{tabular}{c||c|c|c}
 Ch     &  OFF & FF & FF and SRC  
\\
\noalign{\hrule}
 $ a_{p} $  & 0.0004 & 0.0101  & 0.0002  \\
 $ b_{p} $  & 0.0126 & 0.0060  & 0.0031  \\
 $ c_{p} $  & 0.0013 & 0.0303  & 0.1022  \\
 $ d_{p} $  & 0.3918 & 0.1789  & 0.0415  \\
 $ e_{p} $  & 0.1132 & 0.0544  & 0.0346  \\
 $ f_{p} $  & 0.0251 & 0.0121  & 0.0093  \\
\noalign{\hrule}
 $ a_{n} $  & 0.0009 & 0.0202  & 0.0003  \\
 $ b_{n} $  & 0.0251 & 0.0121  & 0.0063  \\
 $ f_{n} $  & 0.0502 & 0.0242  & 0.0186  \\
\noalign{\hrule}
   $\Gamma_{p}   $ & 0.544 & 0.291 & 0.191   \\
   $\Gamma_{n}   $ & 0.076 & 0.056 & 0.025   \\
   $\Gamma_{nm}  $ & 0.621 & 0.348 & 0.216   \\
   $ R_{np}      $ & 0.140 & 0.193 & 0.130   
\end{tabular}
\end{center}
\end{table}

  First we study  the OPE mechanism.
  Table \ref{tab:operesults} shows the partial decay rates
  obtained for the one pion exchange potential.
  All the decay rates are written in the
  unit of $\Gamma_{\Lambda}$, the free $\Lambda$ decay rate.
  We list three sets of results.
  The values listed under ``OFF'' are the results
  when we omit the form factor and the short range correlation
  into account.
  The values listed under ``FF'' are the result
  when we take only the form factor into account.
  The values listed under ``FF and SRC''
  are the final result with  both the form factor
  and the short range correlation taken into account.
  Because $\Delta I=1/2$ is assumed for the weak $\Lambda N \pi$ vertex,
  these partial decay rates satisfy
  the isospin relation $a_n/a_p = b_n/b_p = f_n/f_p= 2$.

  In the ``OFF'' and ``FF'' cases, the channel $d_p$ is dominant.
  This comes from the tensor part of the transition potential,
  which is enhanced due to the large relative momentum in the final state.
  One can see that the form factor reduces most of the
  partial decay rates significantly.
  In the ``FF and SRC'' case,
  the channel $c_p$ becomes large and is dominant.
  This comes also from  the tensor part of the transition potential
  together with the tensor part of the final state interaction.
  In the OPE mechanism, the channel $e_p$ has also a large rate,
  while the rates in the channels $a$ and $b$ are very small.

  Table \ref{tab:operesults} shows calculated 
  nonmesonic decay rates of $^5_{\Lambda}\mbox{He}$ in OPE.
  We list the proton induced decay rate ($\Gamma_p$),
  the neutron induced decay rate ($\Gamma_n$),
  the total decay rate ($\Gamma_{nm}=\Gamma_p+\Gamma_n$),
  and the $n$-$p$ ratio ($ R_{np} = \Gamma_n / \Gamma_p$).
  The experimental data are listed in Table \ref{tab:he5lam}.
  One can see that the calculated $\Gamma_{p}$ is in good agreement
  with experiment, while the calculated $\Gamma_{n}$ is much
  smaller than experiment.
  The $\Gamma_p$ is dominated by large contribution of the
  channels $c$ and $d$,
  which vanish in the neutron induced decay.
  Thus the calculated $n$-$p$ ratio is much smaller than the
experimental one.

\begin{table}[t]
\caption{ Calculated nonmesonic decay rates of $^5_{\Lambda}\mbox{He}$
          in the direct quark mechanism
          (in the unit of $\Gamma_{\Lambda}$,
            the decay rate of the free $\Lambda$). }
\label{tab:dqresults}
\begin{center}
\begin{tabular}{c||c|c|c|c}
      Ch        & $C=0  $ & $C=0.3$  & $C=0.5$ & $C=0.7$
\\
\noalign{\hrule}
    $ a_{p}  $  & 0.0130 & 0.0150 & 0.0167 & 0.0185  \\
    $ b_{p}  $  & 0.0127 & 0.0120 & 0.0113 & 0.0105  \\
    $ c_{p}  $  & 0.0968 & 0.0690 & 0.0548 & 0.0440  \\
    $ d_{p}  $  & 0      & 0      & 0      & 0       \\
    $ e_{p}  $  & 0.0056 & 0.0061 & 0.0064 & 0.0067  \\
    $ f_{p}  $  & 0.0345 & 0.0353 & 0.0353 & 0.0352  \\
\noalign{\hrule}
    $ a_{n}  $  & 0.0727 & 0.0516 & 0.0407 & 0.0322  \\
    $ b_{n}  $  & 0.0059 & 0.0066 & 0.0069 & 0.0073  \\
    $ f_{n}  $  & 0.0622 & 0.0642 & 0.0648 & 0.0650  \\
\noalign{\hrule}
    $\Gamma_{p}   $ & 0.163 & 0.137 &  0.125 & 0.115 \\
    $\Gamma_{n}   $ & 0.141 & 0.122 &  0.112 & 0.104 \\
    $\Gamma_{nm}  $ & 0.304 & 0.260 &  0.237 & 0.219 \\
    $ R_{np}      $ & 0.865 & 0.889 &  0.903 & 0.910 
\end{tabular}
\end{center}
\end{table}

  We turn to the DQ mechanism.
  Table \ref{tab:dqresults} shows the partial decay rates
  when we employ the direct quark induced potential.
  We list four sets of result which are calculated
  with different short range correlations,
  while we assume  $r_i=r_f$ and $C_i=C_f=C$ for simplicity.
  The parameter $r_0$ is fixed as 0.5 fm,
  which is equal to the Gaussian parameter $b$ in the quark model.
  The partial decay rate for channel $d_p$ is zero,
  because the direct quark potential has no $\Delta L = 2$ part
  due to the nonrelativistic truncation at the first order in $p/m$.
  The total decay rates, $\Gamma_p$ $\Gamma_n$ and $\Gamma_{nm}$
  decrease as the short range correlation strength $C_i$ or $C_f$ increases.
  This is natural because the short range correlation
  reduces the inner part of the wave functions
  where the direct quark mechanism is important.

  Compared to the results in Table 2, we find that the overall magnitudes
of the decay rates are comparable to OPE.  The components, however, are
different.  For the proton induced decays, OPE is dominated by the
$I=0$ final states, $c_p$, $d_p$, and $e_p$, while the DQ gives a
large decay rates to the $I=1$ final states, $a_p$, $b_p$, and $f_p$.
The neutron induced decays go only to the $I=1$ states,
$a_n$, $b_n$, and $f_n$, and thus the DQ gives much larger
contribution than OPE.  It should also be stressed that
the $J=0$ channels play major roles in DQ, while the OPE is
dominated by $J=1$.  It is important to note that these
general features can be tested in the decays of the s-shell
hypernuclei.  For instance, we will see that DQ must be dominant in the
decay of $^4_{\Lambda}$H because the $I=0$ final states
are prohibited there.

\begin{table}[t]
\caption{  Calculated nonmesonic decay rates of $^5_{\Lambda}\mbox{He}$
           in the direct quark mechanism
           ($C_i=C_f=C=0.5$).
           The partial rates, $c_p$, $d_p$  and $e_p$ do not have
           $\Delta I=3/2$ contribution because the final state has
$I=0$.
          }
\label{tab:deltai}
\begin{center}
\begin{tabular}{c||c|c}
        Ch           &  $\Delta I =1/2$ only & DQ full    \\
\noalign{\hrule}
       $ a_{p}   $   &  0.0118  & 0.0167 \\
       $ b_{p}   $   &  0.0001  & 0.0113 \\
       $ f_{p}   $   &  0.0334  & 0.0353 \\
\noalign{\hrule}
       $ a_{n}   $   &  0.0237  & 0.0407 \\
       $ b_{n}   $   &  0.0003  & 0.0069 \\
       $ f_{n}   $   &  0.0668  & 0.0648
\end{tabular}
\end{center}
\end{table}

  Next we investigate the $\Delta I$ property of the direct quark mechanism.
  In the following, we employ $C_i=C_f=C=0.5$.
  Table \ref{tab:deltai} shows the effects of the $\Delta I=3/2$
  part of the potential.
  The values listed under ``DQ full'' are the results
  when we employ the full direct quark induced potential.
  While the values listed under ``$ \Delta I=1/2$ only'' are the results
  when we omit the $\Delta I=3/2$ part.
  One sees that the channels $a$ and $b$, which are $J=0$ transitions,
  get significant contribution from the $\Delta I=3/2$ part.
  This indicates that $\Delta I =1/2$ rule is broken significantly
  in the non-mesonic decays of light hypernuclei.

  The large $\Delta I =3/2$ transition is very interesting.
  It is, in fact, expected naturally in the standard theory
  of the weak interaction.
  But one may wonder whether
  the present quark model with Hamiltonian $H_{eff}^{\Delta S=1}$
  is capable to reproduce the $\Delta I =1/2$ dominance of the
  free hyperon decay.
  In order to check it, we study the free $\Lambda$ decay
  with this effective Hamiltonian \cite{lambda-decay}.
  We find the dominance of quark diagrams with internal
  weak vertex in which $\Delta I =3/2$ amplitudes vanish,
  according to the Pati-Woo theorem which is resulted from
  the color symmetry of the ground state baryons \cite{Pati-Woo}.
  Therefore, the dominant decay of the hyperons go through the $\Delta I=1/2$
  part of the Hamiltonian.
  It is extremely interesting to confirm the strong
  $\Delta I = 3/2$ weak transition.

  The calculated $\Gamma_p$, $\Gamma_n$, $\Gamma_{nm}$, and $R_{np}$ in DQ 
  are shown in Table \ref{tab:dqresults}. 
  The proton induced decay rate $\Gamma_{p}$ in DQ is smaller than that of OPE.
  On the other hand, the neutron induced decay rate
  $\Gamma_{n}$ in DQ is much larger than that of the OPE.
  This is due to large contribution of channel $a_n$ and $f_n$.
  The total nonmesonic decay rate in DQ is roughly equal,
  while the $n$-$p$ ratio in DQ is much larger than that of OPE
  and is about 0.9, which is closer to the experimental data.

\begin{table}[t]
\caption{ Calculated nonmesonic decay rates of $^5_{\Lambda}\mbox{He}$
          (in the unit of $\Gamma_{\Lambda}$). }
\label{tab:he5lam}
\begin{center}
\begin{tabular}{c||c|c|c||c|c}
     Ch    &  OPE only  &  DQ only  &  OPE $+$ DQ
           &  EXP \cite{Szymanski}  &  EXP \cite{KishiWein} \\
\noalign{\hrule}
  $ a_{p}   $   & 0.0002  & 0.0167 &  0.0188 &  & \\
  $ b_{p}   $   & 0.0031  & 0.0113 &  0.0026 &  & \\
  $ c_{p}   $   & 0.1022  & 0.0548 &  0.2612 &  & \\
  $ d_{p}   $   & 0.0415  & 0      &  0.0415 &  & \\
  $ e_{p}   $   & 0.0346  & 0.0064 &  0.0207 &  & \\
  $ f_{p}   $   & 0.0093  & 0.0353 &  0.0763 &  & \\
\noalign{\hrule}
  $ a_{n}   $   & 0.0003  & 0.0407 &  0.0356 &  & \\
  $ b_{n}   $   & 0.0063  & 0.0069 &  0.0264 &  & \\
  $ f_{n}   $   & 0.0185  & 0.0648 &  0.1437 &  & \\
\noalign{\hrule}
  $\Gamma_{p}   $ & 0.191 & 0.125 & 0.421 & 0.21 $\pm$ 0.07
                                             &  0.17 $\pm$ 0.04 \\
  $\Gamma_{n}   $ & 0.025 & 0.112 & 0.206 & 0.20 $\pm$ 0.11
                                             &  0.33 $\pm$ 0.04 \\
  $\Gamma_{nm}  $ & 0.216 & 0.237 & 0.627 & 0.41 $\pm$ 0.14
                                             &  0.50 $\pm$ 0.07 \\
  $ R_{np}      $ & 0.132 & 0.903 & 0.489 & 0.93 $\pm$ 0.55
                                             &  1.97 $\pm$ 0.67
\end{tabular}
\end{center}
\end{table}

 As was argued in the section 2, the final results are
given by the superposition of the OPE and DQ processes.
Table \ref{tab:he5lam} summarizes the results given by the sum of the
two transition potentials.  
We find that the neutron induced decay rate, $\Gamma_n$,
is significantly enhanced from OPE and 
becomes consistent with experimental data.
On the other hand, the combined result 
overestimates the proton induced decay rate, $\Gamma_p$.
Thus the total decay rate, $\Gamma_{nm}  $, is slightly overestimated.
The $n$-$p$ ratio, $ R_{np}$, is predicted in between the values for 
OPE and DQ, while the experimental data suggest larger values.
%In all, the results
%agree reasonably well with the observed values within the error
%except for $\Gamma_p$.

\subsection{ $^4_{\Lambda}\mbox{He}$ }

\begin{table}[t]
\caption{ Calculated nonmesonic decay rates of $^4_{\Lambda}\mbox{He}$
          (in the unit of $\Gamma_{\Lambda}$). }
\label{tab:he4lam}
\begin{center}
\begin{tabular}{c||c|c|c||c}
  Ch     & OPE only & DQ only &  OPE$+$DQ  & EXP \cite{Val} \\
\noalign{\hrule}
  $ a_{p}   $ & 0.0001  & 0.0183  & 0.0214 &    \\
  $ b_{p}   $ & 0.0021  & 0.0087  & 0.0023 &    \\
  $ c_{p}   $ & 0.0818  & 0.0004  & 0.0884 &    \\
  $ d_{p}   $ & 0.0321  & 0       & 0.0321 &    \\
  $ e_{p}   $ & 0.0224  & 0.0048  & 0.0140 &    \\
  $ f_{p}   $ & 0.0065  & 0.0275  & 0.0567 &    \\
\noalign{\hrule}
  $ a_{n}   $ & 0.0005  & 0.0013  & 0.0004 &    \\
  $ b_{n}   $ & 0.0083  & 0.0108  & 0.0380 &    \\
  $ f_{n}   $ & 0       & 0       & 0      &    \\
\noalign{\hrule}
  $\Gamma_{p}   $ & 0.145 & 0.060 & 0.214 &  0.15 $\pm$ 0.02 \\
  $\Gamma_{n}   $ & 0.009 & 0.012 & 0.038 &  0.04 $\pm$ 0.02 \\
  $\Gamma_{nm}  $ & 0.154 & 0.072 & 0.253 &  0.19 $\pm$ 0.04 \\
  $ R_{np}      $ & 0.061 & 0.202 & 0.178 &  0.27 $\pm$ 0.14 \\
\end{tabular}
\end{center}
\end{table}

  In the $^4_{\Lambda}\mbox{He}$, there are two $\Lambda p$ bonds
  and one $\Lambda n$ bond.
  The $\Lambda n$ pair is in the spin S=0 state
  so that the spin of $^4_{\Lambda}\mbox{He}$ is equal to zero.
  Thus there is no contribution from the channel $f_n$.
  The the partial decay rates of are given by
%
%\eq
%  \Gamma_{nm}(^4_{\Lambda}\mbox{He})
%   =
%          \Gamma_a^p + \Gamma_b^p
%       +  \Gamma_c^p + \Gamma_d^p + \Gamma_e^p + \Gamma_f^p
%       +  \Gamma_a^n + \Gamma_b^n
%\eq
%  where
%
\eq
  \Gamma_{a,b}^p
   &=&
   2 \frac14
   \disint \frac{d^3 \mvec k}{(2\pi)^3}
   \disint \frac{d^3 \mvec K}{(2\pi)^3}
   (2 \pi)\delta(E.C.)
   \left|
      \bra {NN} |V_{a,b}^p|{\Lambda N} \ket
   \right|^2
   \\
   \Gamma_{c,d,e,f}^p
   &=&
   2  \frac34
   \disint \frac{d^3 \mvec k}{(2\pi)^3}
   \disint \frac{d^3 \mvec K}{(2\pi)^3}
   (2 \pi)\delta(E.C.)
   \left|
      \bra {NN} |V_{c,d,e,f}^p|{\Lambda N} \ket
   \right|^2
  \\
  \Gamma_{a,b}^n
  &=&
   \disint \frac{d^3 \mvec k}{(2\pi)^3}
   \disint \frac{d^3 \mvec K}{(2\pi)^3}
   (2 \pi)\delta(E.C.)
   \left|
      \bra {NN} |V_{a,b}^n|{\Lambda N} \ket
   \right|^2
 \ .
\eq

  Table \ref{tab:he4lam} shows our results and the experimental data.
  The results of OPE are qualitatively the same to
  $^5_{\Lambda}\mbox{He}$,
  but are reduced to 80 \% or less.
  On the other hand, the results of DQ have qualitative differences.
  The rates in the channels  $c_p$ and $a_n$ are very small,
  which are large in the $^5_{\Lambda}\mbox{He}$ case.
  This indicates that the DQ contribution is sensitive to
  the wave function of the initial $\Lambda N$ system.
  Again the full calculation of OPE+DQ provides a good
  agreement to the experimental data except that $\Gamma_p$ is
  again slightly overestimated.
  In our result, $\Gamma_n$ is dominated by the channel $b$.

\subsection{ $^4_{\Lambda}\mbox{H}$ }

\begin{table}[t]
\caption{ Calculated nonmesonic decay rates of $^4_{\Lambda}\mbox{H}$
          (in the unit of $\Gamma_{\Lambda}$). }
\label{tab:h4lam}
\begin{center}
\begin{tabular}{c||c|c|c||c}
  Ch     & OPE only & DQ only & OPE$+$DQ  & EXP \cite{Outa} \\
\noalign{\hrule}
  $ a_{p}   $   & 0.0003  & 0.0366 & 0.0429 &  \\
  $ b_{p}   $   & 0.0041  & 0.0174 & 0.0046 &  \\
\noalign{\hrule}
  $ a_{n}   $   & 0.0003  & 0.0006 & 0.0002 &  \\
  $ b_{n}   $   & 0.0041  & 0.0054 & 0.0190 &  \\
  $ f_{n}   $   & 0.0130  & 0.0504 & 0.0107 &  \\
\noalign{\hrule}
  $\Gamma_{p}   $ & 0.004 & 0.054 & 0.047 &  \\
  $\Gamma_{n}   $ & 0.017 & 0.057 & 0.126 &  \\
  $\Gamma_{nm}  $ & 0.022 & 0.110 & 0.174 &  0.15 $\pm$ 0.13\\
  $ R_{np}      $ & 3.952 & 1.048 & 2.660 &  \\
\end{tabular}
\end{center}
\end{table}

  In $^4_{\Lambda}\mbox{H}$, there are one $\Lambda p$ bond
  and two $\Lambda n$ bonds.
  The $\Lambda p$ pair is in the spin S=0 state
  so that the spin of $^4_{\Lambda}\mbox{He}$ is equal to zero.
  Thus there are no contributions from the channels $c_p$ though $f_p$.
  The the partial decay rates are given by
%
%\eq
%  \Gamma_{nm}(^4_{\Lambda}\mbox{H})
%    =   \Gamma_a^p + \Gamma_b^p  +  \Gamma_a^n + \Gamma_b^n +
%\Gamma_f^n
%\eq
% where
%
\eq
   \Gamma_{a,b}^p
   &=&
   \disint \frac{d^3 \mvec k}{(2\pi)^3}
   \disint \frac{d^3 \mvec K}{(2\pi)^3}
   (2 \pi)\delta(E.C.)
   \left|
      \bra {NN} |V_{a,b}^p|{\Lambda N} \ket
   \right|^2
  \\
  \Gamma_{a,b}^n
  &=&
   2 \frac14
   \disint \frac{d^3 \mvec k}{(2\pi)^3}
   \disint \frac{d^3 \mvec K}{(2\pi)^3}
   (2 \pi)\delta(E.C.)
   \left|
      \bra {NN} |V_{a,b}^n| {\Lambda N} \ket
   \right|^2
  \\
  \Gamma_{f}^n
  &=&
   2 \frac34
   \disint \frac{d^3 \mvec k}{(2\pi)^3}
   \disint \frac{d^3 \mvec K}{(2\pi)^3}
   (2 \pi)\delta(E.C.)
   \left|
      \bra {NN} |V_{f}^n| {\Lambda N} \ket
   \right|^2
 \ .
\eq

  Table \ref{tab:h4lam} shows our results.
  In our model, each partial decay rate is the same as
  that of $^4_{\Lambda}\mbox{He}$ except for the number of bonds
  and the spin average factor.
  The effect of the interference of OPE and DQ is large in the channel
  $f_n$,
  which does not appear in the $^4_{\Lambda}\mbox{He}$ decay.
  At present, the experimental data are very limited.
  Only the total nonmesonic decay rate is barely known.
  Our prediction agrees with it.
  The result also suggests that the
  neutron induced decay is stronger than the proton.
  Thus we expect a large $n$-$p$ ratio for
  $^4_{\Lambda}\mbox{H}$.
  We anticipate new good experimental data for
  the nonmesonic decay of $^4_{\Lambda}\mbox{H}$.

\subsection{Other contributions}

  So far we considered only the one-pion exchange and the direct quark
  transitions. Several other possibilities are considered in the
  following.
  First we discuss the effect of the heavy meson exchanges.
  Ramos \etal~ calculated the $\Lambda N \to NN$
  transition in a full one-boson-exchange mechanism \cite{Ramos2}.
  They include the other pseudosclar mesons, $\eta$ and $K$,
  as well as the vector mesons, $\rho, \omega$ and $K^*$.
  In constructing the transition potential induced by the
  exchange of the heavy mesons, the Nijmegen or the J\"ulich strong
  vertices,
  $H_{N N \eta}, H_{N N K}, H_{N \rho}, H_{N N \omega}$ and $H_{N N K^*}$,
  are employed.
  On the other hand, the weak vertices,
  $H_{\Lambda N \eta}, H_{\Lambda N K}, H_{\Lambda N \rho},
  H_{\Lambda N \omega}$ and $H_{\Lambda N K^*}$, which cannot
  be determined from the hyperon-decay experiment,
  are determined by the $SU(6)$ symmetry and the soft meson theorem
  for the PV vertices and pole model for the PC vertices \cite{torre}.
  The nonmesonic weak decay of $^{13}_{\Lambda}C$
  is studied in the shell model framework.
  The result shows that the combined $\pi+\rho$ exchange
  predicts a very similar total decay rate to the OPE one.
  It is found that the total rate is reduced by more than 40\%
  when $K$ exchange is added.
  It is also found that $K^*$ exchange compensates the $K$ exchange
  and that the total rate for the combined $\pi + \rho + K + K^*$
  is only 10 \% smaller than the OPE only.
  The contribution of $\eta$ and $\omega$
  are very small and tend to cancel each other.
  Therefore the total rate for the combined
  $\pi + \rho + K + K^* + \eta + \omega$ is similar to the OPE only.
  They found that the addition of $K$ exchange
  reduces the $n$-$p$ ratio considerably
  while the addition of other mesons does not change the result much.
  The final $n$-$p$ ratio is smaller than that of OPE,
  which is much smaller than the experimental data.

  Shmatikov studied the $2\pi$ exchange contribution.
  It is indicated that the diagrams with $\Sigma N$ and $NN$
  intermediate states cancel each other and the net
  effect contributes only to the $J=0$ amplitudes \cite{Shmatikov2}.

  Recently, 
Itonaga \etal~ \cite{Itonaga} have studied the 
$2\pi/\rho$ and $2\pi/\sigma$ 
exchange weak potentials in which $2\pi$ are coupled to $\rho$ and $\sigma$, 
respectively, in the exchange process. The weak vertices $H_{\Lambda N \pi}$ 
and $H_{\Sigma N \pi}$ are empirically known from the pionic decay of 
$\Lambda$ and $\Sigma$ hyperons and the Lee-Sugawara relations. The strong 
coupling constants are determined so that the corresponding strong potential 
versions of the $2\pi/\rho$ and $2\pi/\sigma$ exchanges can simulate the OBE 
potentials due to $\rho$ and $\sigma$ exchange, respectively. There is no 
phase ambiguity between OPE and $2\pi/\rho$ and $2\pi/\sigma$ exchange 
weak potentials. 
The $2\pi/\rho$ exchange potential has a tensor part whose sign is opposite 
to that of the OPE one and thus it tends to weaken the 
tensor interaction of the latter potential. The $2\pi/\sigma$ exchange is 
of central type typical from its character. These potentials are applied 
to the non-mesonic decays of typical hypernuclei raging from s-shell to 
medium-heavy systems. The addition of $2\pi/\rho$ exchange decreases 
the non-mesonic decay rates by 5-10 \% with respect to the OPE estimates. 
The inclusion of both $2\pi/\rho$ and $2\pi/\sigma$ exchanges leads to 
increase the decay rates of about 5-15 \% with respect to the OPE estimate. 
Calculated non-mesonic decay rates, together with minor contributions from 
the pionic decay channels, are generally consistent with the lifetimes 
measured for p- and sd-shell hypernuclei. As for the n/p ratios, the addition 
of the $2\pi/\rho$ and $2\pi/\sigma$ exchanges does not cause much 
improvement and the theoretical values are still far from the experimental 
data.

  In all, the meson exchange contributions other than OPE seem
  to be less important than DQ.
  Therefore describing the $\Lambda N \to NN$ transition by
  the sum of OPE and DQ, seems to be reasonable.

\section{Conclusion}

  It is shown that the one pion exchange mechanism is significant
  to the nonmesonic decay of light s-shell hypernuclei
  but is not able to reproduce some of the experimental data.
  Our result shows that the direct quark processes
  in $\Lambda N \to NN$ provides as large contribution as
  the OPE.
  We also show that the DQ contribution is qualitatively
  different from the one pion exchange mechanism.
  The DQ contribution is dominant in some channels, such as
  the proton induced $J=0$ channels, $a_p$ and $b_p$.
  It is also shown that the direct quark mechanism causes
  a large $\Delta I =3/2$ transition in several channels.
  Our results are qualitatively consistent with those of
  Maltman and Shmatikov ~\cite{Shmatikov}, although the calculated
  amplitudes have quantitative differences.

  In this paper, we have determined the relative phase of OPE and DQ
  by using the soft pion relation.
  Thus we superpose these two mechanisms without ambiguity.
  We reproduce the present data for
  $^5_{\Lambda}\mbox{He}$, $^4_{\Lambda}\mbox{He}$ and
$^4_{\Lambda}\mbox{H}$
  fairly well by superposing OPE and DQ.
  Some predictions are made for the nonmesonic decays of
  $^4_{\Lambda}\mbox{H}$, to which experimental data are very limited.
  It should be stressed that
  the ratios of the partial rates for various channels
  are extremely useful in distinguishing different mechanisms of the
transition.
  Further experimental studies are most desirable.

  There are a number of remaining problems.
  In the present analysis, we have not considered so far the second
order processes with 
  $\Sigma-N$ intermediate states induced by the pion (meson) 
  and/or quark exchanges.
  The short range part of weak transition potential for $\Sigma N \to
NN$
  was also computed in the same direct quark mechanism \cite{OkaInoue2}.
  It is found that the mixing of this potential does not
  change the main feature of present DQ potential,
  though its contribution depend on the probability of the $\Sigma$
mixture and
  is not negligible quantitatively.

  For hypernuclei other than the s-shell systems,
  we need a realistic calculation combined with
  the nuclear structure analysis.
  Further study is needed.

% write \refPR for the Phys Rev formats for the reference
\newif\ifrefphysrev
\def\refPR{\refphysrevtrue
           \typeout{** Reference: Phys Rev format (substyle:Sachiko)}}
\def\refNP{\refphysrevfalse
           \typeout{** Reference: Nucl Phys format (substyle:Sachiko)}}
\refphysrevfalse

%
% References
%
\def \vol(#1,#2,#3){\ifrefphysrev{{\bf {#1}},
{#3} (19{#2})}\else{{{\bf {#1}}(19{#2}){#3}}}\fi}
\def \NP(#1,#2,#3){Nucl.\ Phys.\          \vol(#1,#2,#3)}
\def \PL(#1,#2,#3){Phys.\ Lett.\          \vol(#1,#2,#3)}
\def \PRL(#1,#2,#3){Phys.\ Rev.\ Lett.\   \vol(#1,#2,#3)}
\def \PRp(#1,#2,#3){Phys.\ Rep.\          \vol(#1,#2,#3)}
\def \PR(#1,#2,#3){Phys.\ Rev.\           \vol(#1,#2,#3)}
\def \PTP(#1,#2,#3){Prog.\ Theor.\ Phys.\ \vol(#1,#2,#3)}
\def \ibid(#1,#2,#3){{\it ibid.}\         \vol(#1,#2,#3)}

\end{document}